# A Hybrid Algorithm to Enhance Wireless Sensor Networks security on the IoT

*Ntebatseng MAHLAKE[1], Topside E. MATHONSI[2], Tonderai MUCHENJE[3], Deon DU PLESSIS[4]*

[1,2] Tshwane University of Technology, Private Bag X680, Pretoria 0001, South Africa

[1]Tel: +27795751544, Email: ntebamahlake@gmail.com
[2]Tel: +27123829608, Email: mathonsite@tut.ac.za

[3]Tel: +27 83 797 4472, Email: MuchenjeT@tut.ac.za

[4]Tel: +27123829608, Email: DuPlessisDP@tut.ac.za

**Abstract:** The Internet of Things (IoT) is a futuristic technology that promises to connect tons of devices via the internet. As more individuals connect to the internet, it is believed that communication will generate mountains of data. IoT is currently leveraging Wireless Sensor Networks (WSNs) to collect, monitor, and transmit data and sensitive data across wireless networks using sensor nodes. WSNs encounter a variety of threats posed by attackers, including unauthorized access and data security. Especially in the context of the Internet of Things, where small embedded devices with limited computational capabilities, such as sensor nodes, are expected to connect to a larger network. As a result, WSNs are vulnerable to a variety of attacks. Furthermore, implementing security is time-consuming and selective, as traditional security algorithms degrade network performance due to their computational complexity and inherent delays. This paper describes an encryption algorithm that combines the Secure IoT (SIT) algorithm with the Security Protocols for Sensor Networks (SPINS) security protocol to create the Lightweight Security Algorithm (LSA), which addresses data security concerns while reducing power consumption in WSNs without sacrificing performance.

## I. INTRODUCTION

The IoT paradigm has the potential to be the most exciting and promising field of technology advancement in the future, changing communication perspectives [1]. The IoT is an innovation in communication science that allows you to connect any electrical gadget to the internet. Wireless Sensor Networks (WSNs) are currently being utilized in IoT for communication between nodes and the internet. Because the internet is a public environment with a high risk of intruder attacks, most internet communication protocols use a resource-intensive high-security authentication technique to start clustering and routing procedures [2].

WSNs and growing IoT technology, in particular, may provide an open path for attackers in application domains where CIA (confidentiality, integrity, and availability) is critical. Furthermore, the recent integration and collaboration of WSNs with IoT will present security challenges and issues [3]. Most of the WSNs deployments in IoT are sometimes deployed in the unattended hostile environment for gathering sensitive data or information. Therefore, data leakage and alteration leads to cracking of privacy and security concerns in this environment. This may also lead to users with unauthorized access to access the network and manipulate the security and secrecy of data. In additional, most of IoT devices such as sensor nodes are traditionally resource-constrained devices due to limitations factors such as power consumption, memory footprint, computing abilities, and speed. This make the complete implementation of a cryptographic algorithms difficult due to its limitation factors. Moreover, it makes WSNs vulnerable to variety of attacks and put the security of data at risk. In additional, sacrificing the security and privacy of user data to an unauthorized user with unauthorized access to critical data and information is not an option. Therefore, a certain level of security is essential to protect user data [1, 4].

WSNs have a variety of security problems and attacks due to their design, so it is often recommended to use a lightweight security scheme to avoid the technical overheads imposed, which do not impact any of the overall preferred network performance [4]. The implementation of cryptographic security is needed as the number of sensitive data and information increase drastically and easily gets manipulated and transmitted. The need for lightweight cryptography is widely mentioned and discussed in much literature, but there is limited information nor definition and solution for it. Lightweight cryptography can be implemented in both software and hardware and also for resource-constrained devices to improve the utilization of resources, computational time, power consumption and security [5].

Computation with cryptographic techniques such as data encryption and authenticating the messages sent and received must be performed to enhance the reliability, security, performance, and privacy aspects of wireless devices. As a result, selecting a lightweight cryptographic algorithm for securing wirelessly driven nodes must be considered. Since the nodes have resource-constrained in terms of processing time, resources, computational capabilities, cost, and performance, symmetric-key cryptography is the most widely used technique in WSNs [4, 6]. In this research paper the Lightweight Security Algorithm (LSA) is the proposed scheme to enhance the security in WSNs in IoT. The proposed algorithm integrates Secure IoT (SIT) with Security Protocols for Sensor Networks (SPINS) security protocol is recommended for better security and power efficiency. SPINS offers Security Network Encryption Protocol (SNEP) and the "micro" version of the Timed, Efficient, Streaming, Loss-tolerant Authentication Protocol (µTESLA) protocols as the basic block components for key distribution in sensor networks.

The core SPINS block components, in particular, describe basic primitives for ensuring confidentiality, authentication, data integrity, and weak message freshness [12]. It's designed to address problems such as data security and mitigate different attacks, resource utilization, and power consumption. By ensuring that only fresh and new data is transmitted and that no old data is replayed or retransmitted. Therefore, this research paper takes a proactive approach to reduce the number of attacks, by implementing the strong lightweight security encryption algorithm. The remainder of this research paper is structured as follows: In Section II, this research paper presents the related work. In Section III, this research paper presents the proposed LSA algorithm, and the conclusion and future work is in Section IV.

## II. RELATED WORK

This section summarizes some of the related work on data security and detecting possible attacks in WSNs in IoT that has been done over the years. Many researchers have contributed and provided solutions in this area.

Guo et al, [7] proposed algorithm called Light Encryption Device (LED) is a lightweight block cipher for WSNs security. It has a 64-bit block size, a 64/128-bit key size, and four encryption-decryption cycles. It was a hybrid solution made up of AES and PRESENT. They examine the role of an ultra-light (in fact, non-existent) key schedule and the resistance of ciphers, particularly LED, against related-key attacks first. Furthermore, they also derive simple but intriguing AES-like security proofs for LED in the context of related-key or single-key attacks. LED employs AddConstants, SubCells, ShiftRows, and MixColumns serial operations, as well as a light key scheduling algorithm. Despite being a block cipher, this approach is inefficient in terms of energy consumption per bit and is vulnerable to related key attacks.

Ebrahim and Chong [8] proposed a low complexity cryptographic technique for WSNs named Secure Force (SF). The SF algorithm's design provides a low-complexity architecture for WSNs implementation. The algorithm is similar to the proposed, however, this algorithm is not hybrid. The encryption process is made up of only five encryption rounds in order to save energy. It's based on the Feistel structure, and the encryption part may be done using a simple design that just uses fundamental mathematical operations (AND, OR, XOR, XNOR, shifting, swapping) to save energy and improve data security. This is done to produce enough confusion and data diffusion to withstand various forms of attacks. To make the cipher more complicated, different substitution and permutation techniques are used in addition to data confusion and diffusion. The algorithm's security, on the other hand, is based on its mathematical functioning. They add to the cipher's complexity by increasing the level of confusion and data diffusion. Furthermore, the key expansion technique is quite complicated, as it entails a wide range of mathematical operations that take a long time to compute and cost a lot of energy.

Anwar and Maha [9] suggested a lightweight hybrid cryptographic algorithm for WSNs called Advanced Encryption Standard (AES) and Modified Playfair Cipher (AMPC). AMPC was proposed to improve the security of WSNs by using two cryptographic algorithms for data security, AES and modified PlayFair, and a third method, Diffie-Hellman, to protect the key exchange process. There was security was improvement, however, the power consumption was very high.

Rizk and Alkady [10] proposed Two-phase Hybrid Cryptography Algorithm (THCA) for wireless sensor networks to enhance security. THCA is a hybrid that combine Elliptical Curve Cryptography (ECC) and AES to provide encryption. The THCA security algorithm is designed to provide strong security with minimal key maintenance by combining symmetric and asymmetric cryptographic algorithms. Integrity, secrecy, and authenticity are the three cryptographic primitives it guarantees. For authentication, the XOR-DUAL RSA technique was used, and for integrity, the Message Digest-5 (MD5) algorithm was used. In the case of picture encryption, it is resistant to various types of attacks, but it consumes more power.

Prakash et al, [11] developed the Lightweight hybrid cryptography (LWHC) algorithm, which combined the LED and PRESENT Cipher with the SPECK compact key scheduling technique. RECTANGLE S-Box was utilized to make this system faster and more reliable. The LED, PRESENT, and RECTANGLE S-Boxes are used to encrypt the data. The SPECK algorithm, on the other hand, is also utilized for key scheduling. The proposed system encrypts 64 bits of plain data and performs an XOR operation with a 128-bit schedule key using 64 bits of plain data. The LED encryption employed a standard key algorithm that is vulnerable to key assaults. The proposed scheme, on the other hand, uses the 128-bit SPECK key scheduling technique, which makes it lightweight and secure against various key attacks. This advanced system uses a 128-bit key and a 64-bit block plain text XOR to preserve its robustness. The proposed scheme is a lightweight and secure against various key attacks, but not against other security attacks.

Thus, previous research papers primarily focused on checking how different attacks affect communication in WSNs in IoT. However, the proposed algorithm to be implemented in this research paper will lower the level of possible attacks and enhancing security without compromising the performance of the network and also improving the power consumption within WSNs in IoT. The key contribution of this work is to derive a fair packet transit without being attacked by any attack when a packet is distributed from the source node to the destination node. The next section is going to present the proposed algorithm.

## III. THE PROPOSED LSA ALGORITHM

This research paper integrate existing solutions namely SIT and SPINS when designing the proposed LSA algorithm in order to minimize computational complexity, power consumption, and increasing data security during the process of data transmission while lowering the level of possible attacks in WSNs on the IoT. In addition, the algorithm prevents malicious users/nodes from connecting the WSNs during the data transmission. The proposed algorithm is divided into three phases namely: key expansion, key management and encryption process. The proposed algorithm will initiate the process by inputting the data in to network by the user.

The first phase is the key expansion process on the sensor network. The initialization phase generates the keys to be used in the encryption/decryption phase. After the key expansion process then the second phase called key management takes place, where the SPINS security protocol does security checks, such as checking if the key was not exposed to the attackers. The authentication, confidentiality, integrity and data freshness are checked and validated. If the key is exposed to attackers, it is discarded, and the key expansion block generate new key. If the is key is not exposed to unauthorised user or malicious node, then the key management process makes sure that the network has sufficient power to transport data received or obtained on sensor nodes. If there is insufficient power on the network, the packets will be discarded, this will make the network weak and more prone to different attacks. If there is enough power, then the encryption phase take place where the data is being protected from the attacks and ensure the computational complexity and power consumption are on moderate level.

The LSA is a symmetric key algorithm with a 128 bits key and plain content. The encryption technique in symmetric key technique is consists of encryption adjustments, with each round relying on some numerical abilities to create confusion and diffusion. The key is the most important part of the encryption and decryption processes. The entire security of the information is dependent on this key; if the attacker discovers it, the secrecy of the information is compromised. As a result, vital estimates must be taken into account when developing the key's disclosure as difficult as possible. The computations for Feistel-based encryption are made up of a few adjustments, with each round requiring a different key. The proposed algorithm's encryption is made up of five rounds, so we'll need 5 keys for that reason. Increase in the number of rounds ensures better security, but it also results in an increase in the use of required energy. To boost energy consumption, the proposed cryptographic algorithm is designed for only 5 rounds. The key aging process incorporates sophisticated numerical activities, resulting in the creation of 5 distinctive keys from 16 bytes (128 bits) for each round of encryption. Every encryption round has scientific activities that work with 4 bits of data. The proposed algorithm is influenced by SP and Feistel architectural techniques for keeping computational complexity in a moderate level. It was compared to some commonly known cryptographic algorithm such as AES, DES, LED and HIGHT. The proposed algorithm is expected to have minimized power consumption and low level of attack while enhancing data security without compromising the performance of the network in WSNs in IoT.

## IV. CONCLUSION AND FUTURE WORK

WSNs in IoT will soon be a frequent part of our daily lives. Various energy-dependent devices and sensors will be constantly communicating with one another, and their security must not be compromised. Because data security is the most difficult and pressing issue in WSNs in IoT. This study examine various existing data security techniques, as well as the drawbacks and benefits. The conclusion of the research study is, there is need to develop efficient, enhanced, dynamic data security technique which is still a research topic in progress. In this paper we proposed an encryption algorithm to enhance data security and reduce power consumption without compromising the performance of the network in WSNs in IoT namely LSA.  The proposed solution integrates SIT and SPINS algorithm to develop the proposed LSA to detect malicious activities and modification of data with moderate computational complexity, low power consumption and low level of possible attacks while enhancing data security in WSNs in IoT. MATLAB simulation tool will be used to simulate the algorithm in ability to execute the normal tests. In future, the proposed technique will be evaluated using network simulation tool and the results from the simulation will be presented.

## ACKNOWLEDGEMENT

The authors would like to thank the Tshwane University of Technology for financial support. The authors declare that there is no conflict of interest regarding the publication of this paper.